# On the Asymmetry of Vibrational Energy Flow

German Miño-Galaz*[1], Junia Melin[2], Gonzalo Gutierrez[3], Felipe Bravo[1], Valeria Marquez[1] and Fernando Gonzalez-Nilo[1].

[1]Universidad Andres Bello Center for Bioinformatics and Integrative Biology (CBIB), Facultad en Ciencias Biologicas, Santiago, Chile

[2]ASRC Federal Technical Services, 289 Dunlop Blvd, Al 35824, USA

[3]Group of Nanomaterials, Departamento de Física, Facultad de Ciencias, Universidad de Chile, Las Palmeras 3425, Ñuñoa, Santiago, Chile

*corresponding Author: german.mino.galaz@gmail.com

**Abstract**

A simple model to predict the directionality of vibrational energy flow at molecular level is presented. This model is based on a vibrational energy propagation analysis using *ab intio* molecular dynamics and the Fukui function and local softness reactivity indexes derived from DFT. By using this simple conceptual model we are giving a cogent rationale to previous theoretical and experimental reports of asymmetrical vibrational energy diffusion in proteins and energetic materials. We proposed here three basic rules for the Vibrational Energy Relaxation in molecules: (i) the vibrational energy flows form a soft site to a hard site but not in the opposite direction, (ii) when vibrational energy is injected directly to a hard site, it is trapped in this site, and (iii) if the vibrational energy is pump in a polarizable site and two sites with different softness are available, the energy will propagate to the softest one.

## Introduction

The characterization of intramolecular vibrational energy relaxation (IVR) pathways at molecular level is relevant for the proper understanding of numerous chemical phenomenon, including reaction dynamics[1], enzymatic active site energizing[2], allosteric communication[3,4], energy flow in proteins[5], molecular electronics[6], phononics[7] and shock initiation of energetic materials[8].

In previous publications Miño-Galaz studied the asymmetrical vibrational energy flow through hydrogen bonds in the PDZ-2 protein[3] and in α-helices[9] using classical molecular dynamics. In those reports the author concludes that hydrogen bonds can act as vibrational energy diodes, that is, the vibrational energy flow has a preferred directionality. In this letter we are studying the same vibrational rectification effect on different molecular systems using *ab initio* molecular dynamics simulations (AIMD) and quantum chemistry. Furthermore, we proposed a conceptual model to explain the vibrational rectification phenomena based on the Fukui Function *f (r)*, and the local softness *s(r)*, two reactivity indicators defined in the Density Functional Theory (DFT) context.

In order to test our model we are also analyzing a recent publication by *Pein et. al.*[10] where they performed a series of experiments on liquid nitrobenzene to investigate the IVR flow from the phenyl to the nitro group and vice versa. Using IR-Raman spectroscopy the authors can selectively pump energy to different parts of the molecule studying how the vibration is transfer from one region of the molecule to another. They have found that when energy is pumped to the nitro group ***stays*** on the nitro group. Inversely, when energy is input to the phenyl group the vibrational excitation ***does*** travels towards the nitro group. From this novel investigation they conclude that nitrobenzene behave as a vibrational energy diode where the IVR flow is unidirectional. In a later article[11] the authors used the same methodology to study 2-nitrobenzene and 2-fluorobenzene and demonstrate that the vibrational energy flow can be modulated by the $-CH_3$ or $-F$ substituents. The idea of using this phenomenon to construct molecular thermal diodes is certainly very attractive. We are particularly interested in developing new energetic materials where the vibrational energy can be concentrated to create hot spots that would have an impact in the initiation of the energy release[12].

Based on our findings we proposed three basic rules for the vibration energy flow: (i) The vibrational energy flows form a soft site to a hard site but not in the opposite direction, (ii) when vibrational energy is injected directly to a hard site, it is trapped in this site, iii) if the vibrational energy is pump in a polarizable site and two sites with different softness are available, the energy will propagate to the softest one.

## Methods and Computational Details

The simulation of IVR pathways through a hydrogen bond was performed in a selection of four PDZ-2 protein residues (see Fig. 4 in Ref. 9) using AIMD routines implemented in VASP[13,14]. Figure 1 depicts the 3-D working structure containing the four amino acid residues with a total of 72 atoms. In our calculations some of these atoms are fixed (marked with a red) while the perturbation sites are on the ***Amide a*** and ***Amide b*** that belong to the peptide bonds. The structure was confined in a box of 23 x 23 x 28 Å$^3$ where the distance between the images is fixed to 13 Å. The initial setup was equilibrated within the canonical ensemble (NVT) at 10 K for 500 fs. Following this stage an evolution under the microcanonical ensemble (NVE) for 1000 ps was allowed. For the exchange-correlation function the generalized gradient approximation of Perdew, Burke, and Ernzerhof (PBE) was used[13,14]. The core-valence interactions were described by the projector augmented-wave (PAW) potentials[15]. A basis set of plane waves with kinetic energy cutoff of 400 eV were used throughout the calculations[14].

The vibrational perturbations were introduced at the COa and NHb sites by independently stretching the O or H atoms along the bond axis. For the COa bond, the equilibrium distance was determined at 1.241 Å and we are stretching it to 1.32 Å which create an energy pump of 0.216 eV. In the case of NHb, the equilibrium distance is 1.016 Å that we elongate to 1.04 Å, creating an energy pump of 0.290 eV. These two separated setups were allowed to relax in the NVE ensemble for 1000 fs with the same exchange functional and base described above.

The quantum level calculations on N-methyl acetamide, nitrobenzene, 2-nitrobenzene and 2-fluorobenzene[10,11], have been performed at the B3LYP/6-31+g(d,p) level of theory using Gaussian09[16]. The reactivity indicators derived from conceptual DFT and used to generate our model, namely, the Fukui function *f⁻(r)*, and the local softness *s⁻(r)*, have been previously define[17,18] and successfully applied to a number of problems in chemistry[19]. In this letter we are using the condensed to atoms version of these indicators[20]:

$$f_k^- = q_k^{N-1} - q_k^N$$

$$s_k^- = S f_k^-$$

Here the subscript **k** indicates the atomic site, **q** correspond to the atomic charge and **N** is the total number of electrons in the molecule. In the equation for the local softness, **S** represents the total softness of the molecular system[21].

## Results

Our working model is an atomic arrangement for Lys5 – Ile6 ⋯ Asp91 – Glu90 of the PDZ-2 protein shown in Figure 1, along with the perturbation sites. Here, the distances COa/NHa and COb/NHb belong to *Amide a* and *Amide b*, respectively. These amide groups interact through one hydrogen bond. In this computational experiment the vibrational perturbation is introduced by stretching the COa, and the NHb bonds separately. The results of these pump-probe tests in terms of distances and kinetic energies of *Amide a* and *Amide b* are shown in Figure 2. For the sake of comparison we have included a control case where no perturbation was performed. The control case (Plot I) clearly shows that R[CO] and R[NH] distances in both amide groups have very small oscillations, as can be expected for an equilibrium structure. The kinetic energy plots of the unperturbed *Amide a* and *b* (II and III) are showing that for C, N and H atoms the energy oscillation is minimal, below 1.5 kJ/mole, while for O atoms the oscillation can range up to 5kJ/mol.

When the vibrational pump is introduced on *Amide a*, that is, we are stretching R[COa], the oscillation of R[COb] gradually increased as the energy transfer take place (blue line of plot Ia in

Figure 2). This effect can be also followed by changes in the kinetic energy, where perturbations in *Amide a* (IIa) generate a response on *Amide b* by increasing the kinetic energy of all atoms involved, particularly in Cb and Ob (red and blue lines in plot IIIa respectively). On the other hand, when the energy is pump to *Amide b* by stretching R[NHb] from equilibrium to 1.14 Å, the oscillation pattern of this bond (green line in Ib) remain constant while the oscillation of R[COa] and R[NHa] do not seemed to be perturbed. In addition, the kinetic energy plot for *Amide a* in this case (IIb) is very similar to the control case, while the kinetic energy plot for *Amide b* (IIIb) clearly shows how the vibrational energy is trapped in the NHb bond. From these interesting results it is clear that the vibrational energy flow is unidirectional. Furthermore, we can conclude that the vibrational rectification effect is actually generated at the amide bond. In order to prove this hypothesis we studied the local properties of electron density in a smaller system containing this type of bond, namely the N-methyl acetamide.

The Fukui function, $f(r)$, is one of the local indicators most widely used to explain chemical reactivity because of its connection to Frontier Molecular Orbitals FMO theory. In the present analysis we are always referring to the nucleophilic Fukui Function (denoted by the negative superscript) since it is directly related to the HOMO distribution. Using the expression showed in the previous section, we have calculated the nucleophilic Fukui function condensed to atoms ($f_k^-$) describing the amide bond within N-methyl acetamide molecule. Figure 3 shows the HOMO distribution for the equilibrium structure (Fig. 3a) and for the compressed R[CO] and R[NH] (Fig. 3b and 3c respectively). These figures show that the HOMO is drastically reorganized when changing the CO bond distance, but it remains unaltered when the NH bond is perturbed. This qualitative result can be quantified through the condense to atoms Fukui Function values; for the control case the oxygen atom presents the largest $f_k^-$ (Figure 3a), however when compressing the CO bond the nitrogen is activated and the Fukui function increases from 0.062 to 0.295. On the contrary, when shortening NH bond in this molecule the Fukui Function value for C, O, N, H atoms remain almost identical to the control case.

While $f(r)$ is a regioselectivity index, its twin function the local softness, $s^-(r)$, provides information on how the electron density would polarized in response to the vibrational perturbations because the local softness is defined as the projection of $f(r)$ onto the global softness, $S$, of the molecule. In Figure 4 we present the results of scanning this local property, $s_k^-$, along the changes in the interatomic distance between CO and NH bonds while keeping the rest of the atoms fixed. Using the mathematical expression previously described, we have calculated the local softness by simple multiplying $f_k^-$ values to the finite difference approximation global softness ($S = 1/(\varepsilon_{LUMO} - \varepsilon_{HOMO})$).

The scanning along the CO bond, form 1.40 to 1.15 Å, clearly generate responses in the O and N atoms of the molecule (red and green lines in 4a, respectively) while leaving the response of C and H atoms essentially constant. The scanning of the NH bond, form 1.20 to 0.92 Å leaves the response of O, C, N and H atoms constant (Figure 4b). This essentially means that the perturbation of the CO bond is capable of polarize the electron density of the amide but the perturbation in the NH bond does not produce the same response. The effect is similar to the pair site reactivity model to explain the activation of the catalytic site of α-chymotrypsine[22].

Based on our findings, we proposed an operative mechanism for the ***asymmetric vibrational energy flow*** at quantum level: First, let us notice that the CO moiety is the softest site of the amide, thus is easier to polarize its electron density, the NH moiety is the hard site of the amide and thus its electron density would not be easily polarized. Hence, we hypothesize that the perturbation at the soft site of the molecule generates a response of the electron density and since the electron movements are coupled through the bonds, this perturbation propagates towards the other moieties. The changes in the electron density alter the external potential over N and H atoms and they are force to respond. In the other hand, the perturbations done at the hard (or less soft) site of the amide, does not polarized the electron density and therefore does not produced a response in the soft site. The projection of the electrostatic potential upon the electron density for N-methyl acetamide, show in Figure 4c for three selected geometries, reinforces this hypothesis. The compression of the CO bond at a distance R[CO]=1.15 Å alters the charge distribution upon the N atom towards the negative region of the electrostatic potential while compressing the NH bond at a distance R[NH]=0.92 Å does not produce significant changes in any region of molecule, leaving it essentially similar to the equilibrium case.

The conceptual model for vibrational energy flow is now tested on a different set of molecules. Pein el at[10] have experimentally shown that for liquid nitrobenzene at room temperature the vibrational energy transfers from the phenyl moiety to the nitro group with the inverse direction impeded. In order to analyze this system with our model it seems appropriated working with the so-called "group softness". The group softness has proven useful to study other biological systems[22,23] and is simple defined as the linear combination of the local softness of every atomic site that belongs to that functional group, at a fixed geometry. Figure 5a depicts the HOMO distribution for nitrobenzene on the left and the $s_k^-$ for this system on the right. The values after the brackets are the group softness. The softness of the phenyl group is 0.197 eV$^{-1}$ while the nitro group shows a much lower value, 0.004 eV$^{-1}$. The injection of vibrational energy at to the soft portion of the molecule, i.e. the phenyl group enables the polarization of its electron density; hence the electron density of the nitro group is also perturbed. The changes at this site alters the potential that the -NO$_2$ and the whole molecule experience and the vibrational energy is allowed to flow towards the nitro group and the rest of the molecule. The inverse procedure, the pump of energy at the proper frequency on the nitro group (the hard site of the molecule) does not reorganized the electron density and the vibrational perturbation gets trapped in the nitro group.

Further experimental research done by Pein and Dlott[11] has shown that the directionality of vibrational energy flow between the phenyl and nitro group can be modulated introducing a substituent to the nitrobenzene, thus they have studied 2-methyl nitrobenzene and 2-fluoro nitrobenzene. Now, when the vibrational perturbation is over the nitro group, the energy will flow towards the phenyl group and to the rest of the molecule, including the substituent (in both cases). However, if the pump is excerpted at the phenyl group the energy flows preferentially to the substituent (-CH$_3$ or –F). Again, these experimental observations can be rationalized using the Fukui function (here represented by the HOMO plots) and the group softness (Figures 5b and 5c). When adding a substituent to nitrobenzene in the -*ortho* position the HOMO distribution changes and the redistribution affects the $s_k^-$ values. Notice that mayor change involves the disappearance of a node at the C atom, connecting the nitro and phenyl moieties, increasing the polarizability of this site from 0.001 to in nitrobenzene to 0.008 and 0.013 eV$^{-1}$ for the 2-methyl and 2-fluoro nitrobenzene, respectively. This open a channel in which any deformation of the electron density at the nitro group can now be readily sensed by the phenyl ring, and the concomitant vibrational perturbation can be transferred to the rest of the molecule. Interestingly, when the energy pump is excerpted over the phenyl group the preferential pathway of energy flow is towards the -CH$_3$ or -F substituents rather than to the nitro group.

## Conclusions

The agreement between our model of local softness and Pein's experimental findings allow us to propose three rules for the Vibrational Energy Relaxation; (i) The vibrational energy flows form a soft site to a hard site but not in the opposite direction, (ii) when vibrational energy is injected directly to a hard site, it is trapped in this site, iii) if the vibrational energy is pump in a polarizable site and two sites with different softness are available, the energy will propagate to the softest one. Notice that all five systems analyzed here; the four amino acids setup of the PDZ-2 protein, N-methyl acetamide, nitrobenzene, 2-methyl nitrobenzene and 2-fluoro nitrobenzene, consistently obey these rules suggesting the robustness of the model. Using this conceptual model we anticipated engineering molecules with specific directionality of vibrational energy flow that may be used for the design of molecular thermal diodes and low-ignition energetic materials. This new finding can be considered as an extension of the hard-soft acid base (HSAB) principle of Pearson[24]. Thus, we propose the name of vibrational HSAB principle (vHSAB) for it.

**Captions to the figures.**

Figure 1. Setup of a segment of the PDZ-2 protein including residues Lys5 – Ile6 ··· Asp91 – Glu90 interacting by a hydrogen bond. Red dots indicate fixed atoms. See text for details.

Figure 2. Plots of distances R[COa], R[NHa], R[COb] and R[NHb] and kinetic energy per atom in Amide a and b. Sets I, II, II the results when no perturbation is excerpted (control case). Sets Ia, IIa IIIa shows the results when the pump is done in COa. Sets Ib, IIb IIIb shows the results when the pump is done in NHb.

Figure 3. HOMO plots and $f^-_k$ indexes for N-methyl acetamide in three cases: a) Control case (minimum). b) R[CO]=1.15 Å and c) R[NH]=0.92 Å. Indexes for methyls groups are added up.

Figure 4. Plots of $f^-_k$ and $s^-_k$ indexes for dimethyl amide as a response of a (a) stretching at CO moiety and at (b) NH moiety and $s^-_k$ indexes. (c) Electrostatic potential surface projected in a isosurface of electron density (0.02 e/b$^3$) at equilibrium, at R[CO]=1.15 Å and at R[NH]=0.92 Å. Color scale: red, most negative; blue, most positive.

Figure 5. HOMO plots and $s^-_k$ indexes for nitrobenze, 2-methy nitrobenzene, 2-fluro nitrobenzene. Indexes for methyl group in methyl group of 2-methy nitrobenzene, 2-fluro nitrobenzene are added up.

Figure 1.

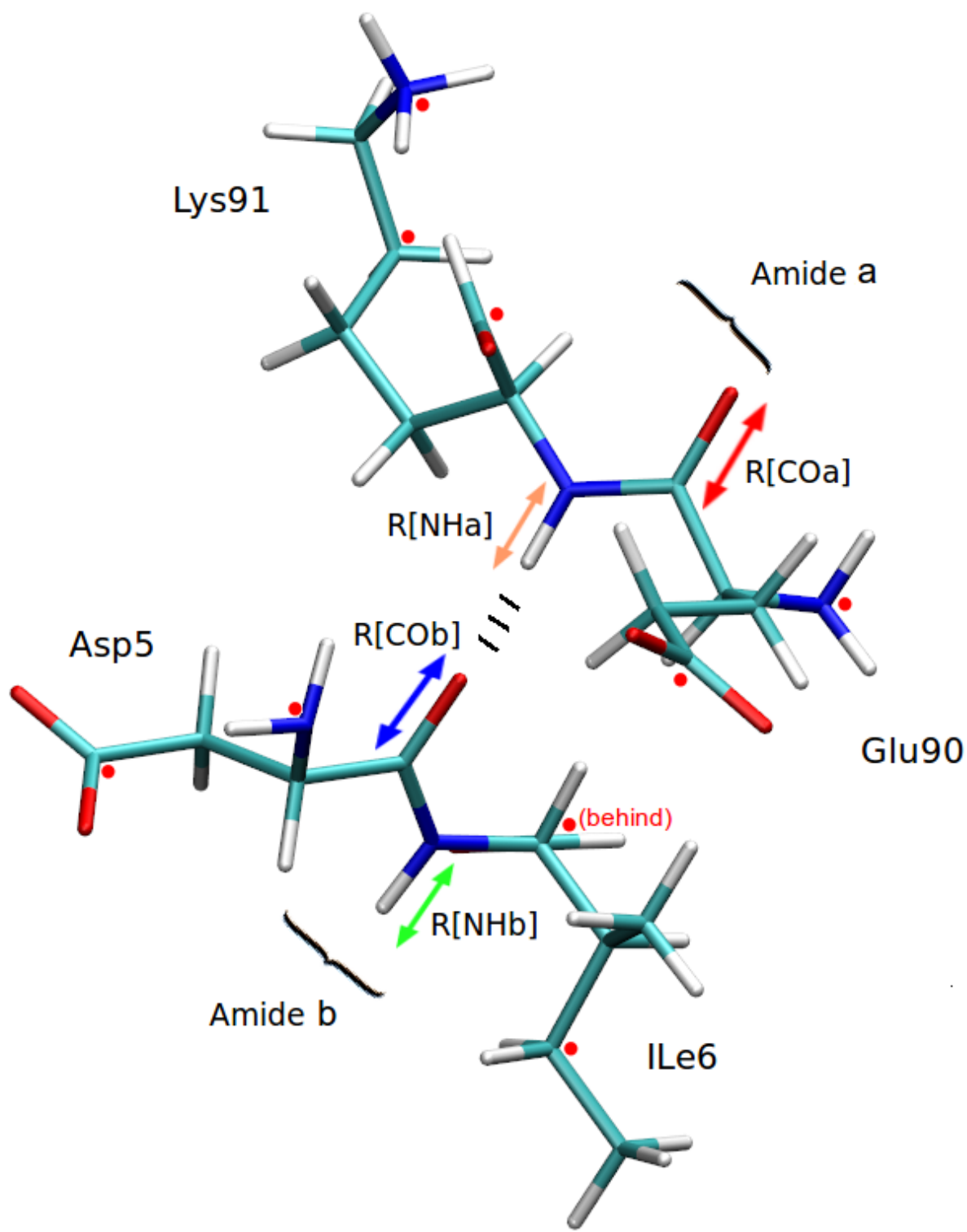

Figure 2.

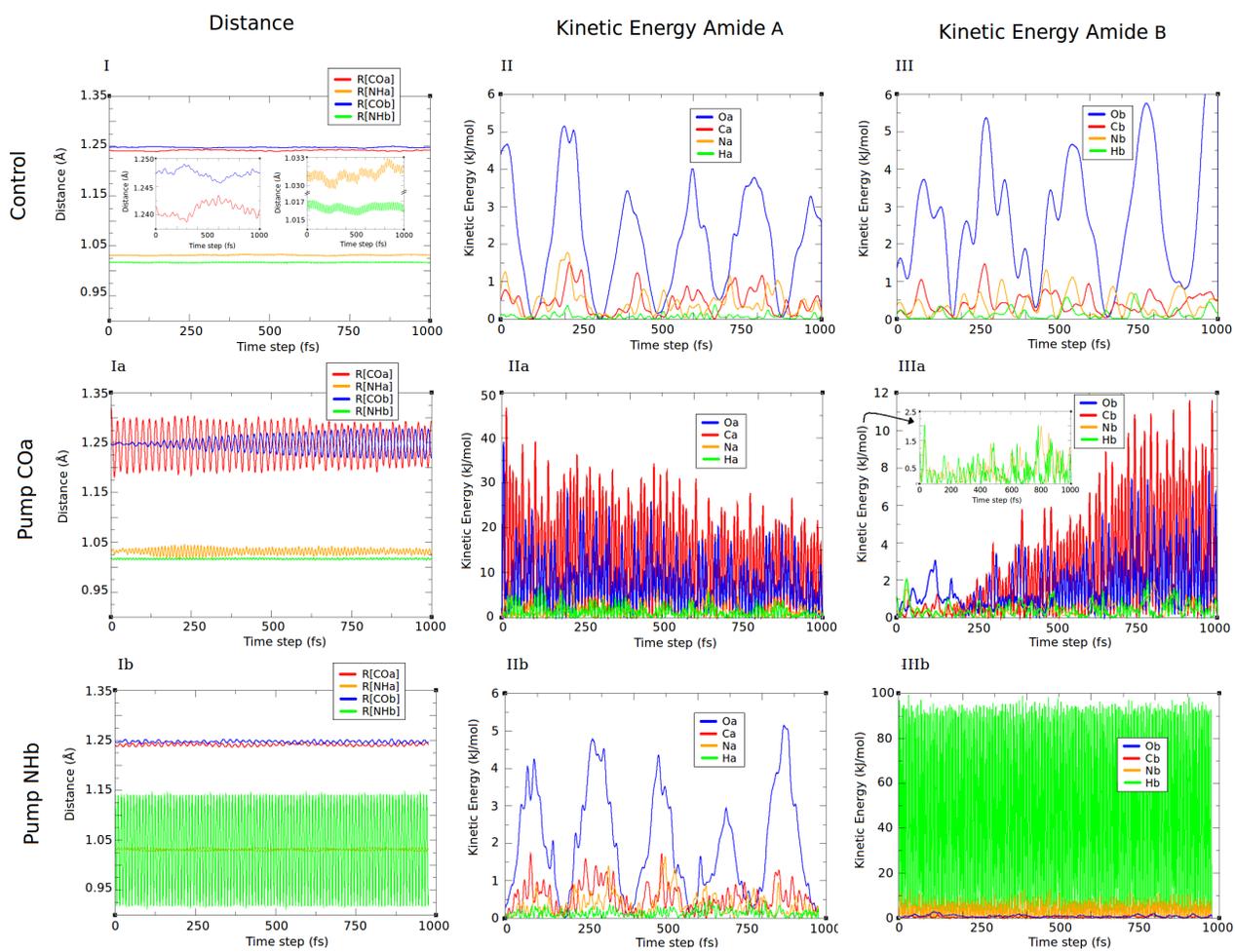

Figure 3.

Homo Plot

a) Control

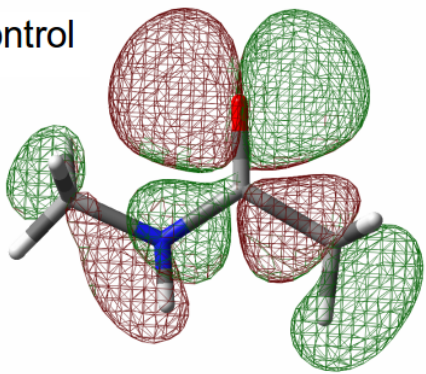
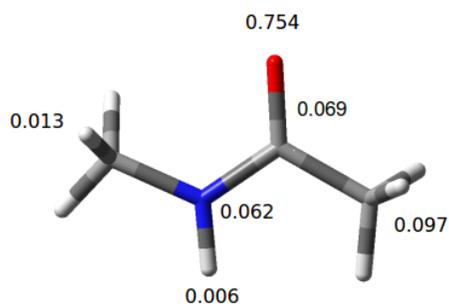

b) R[CO]=1.15 Å

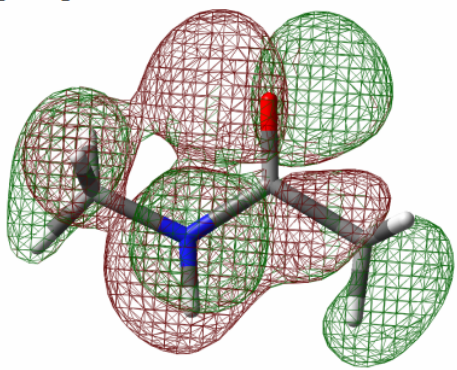
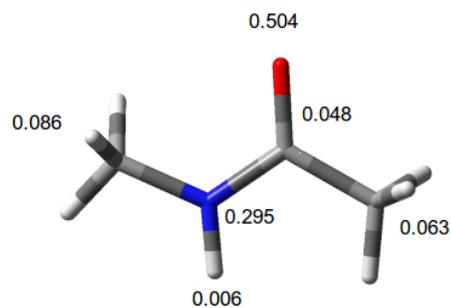

c) R[NH]=0.92 Å

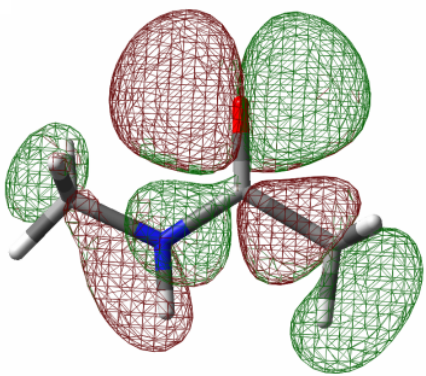
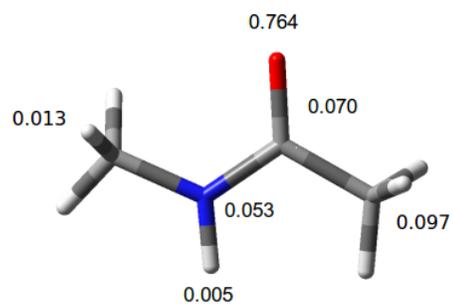

Figure 4.

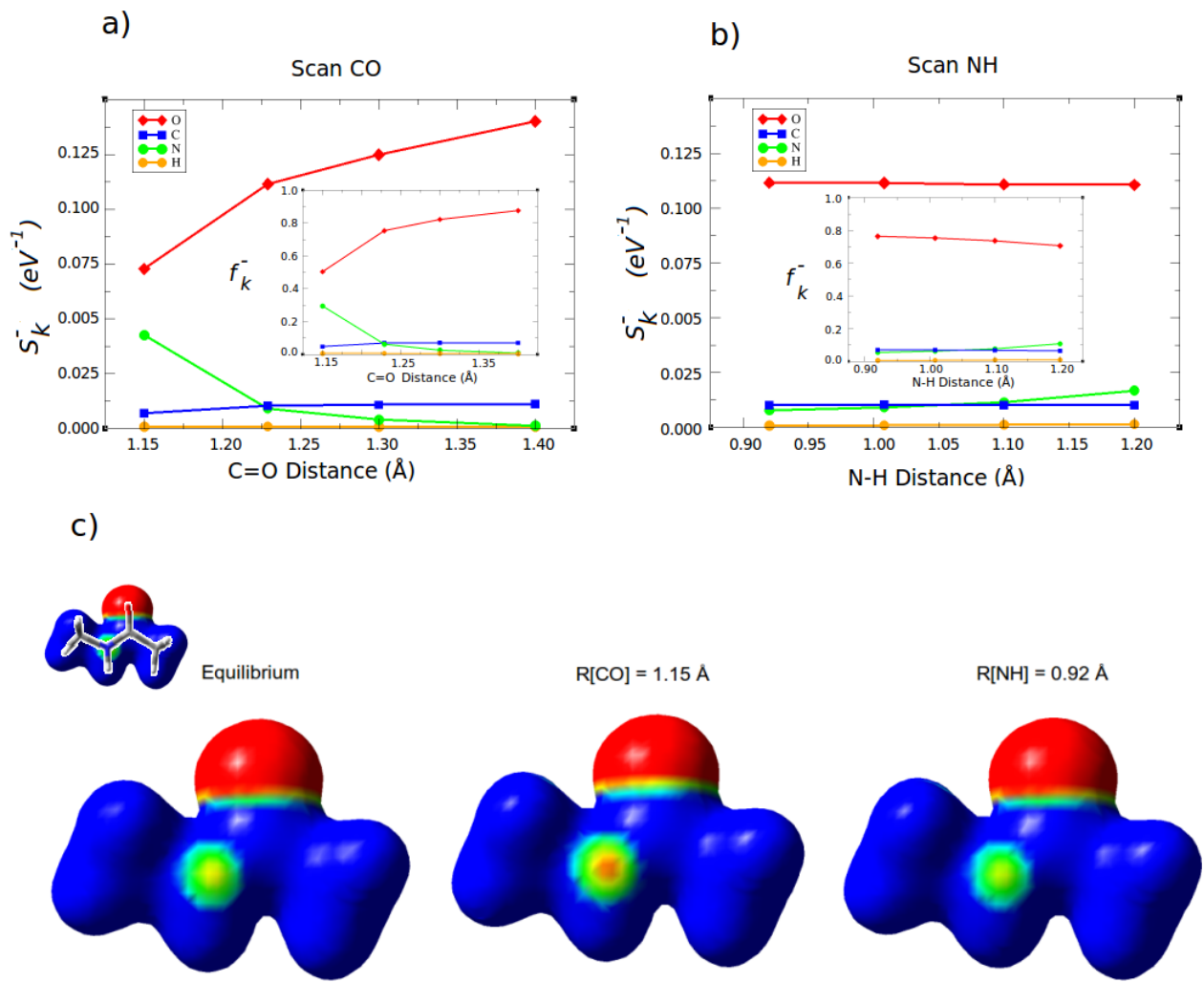

Figure 5.

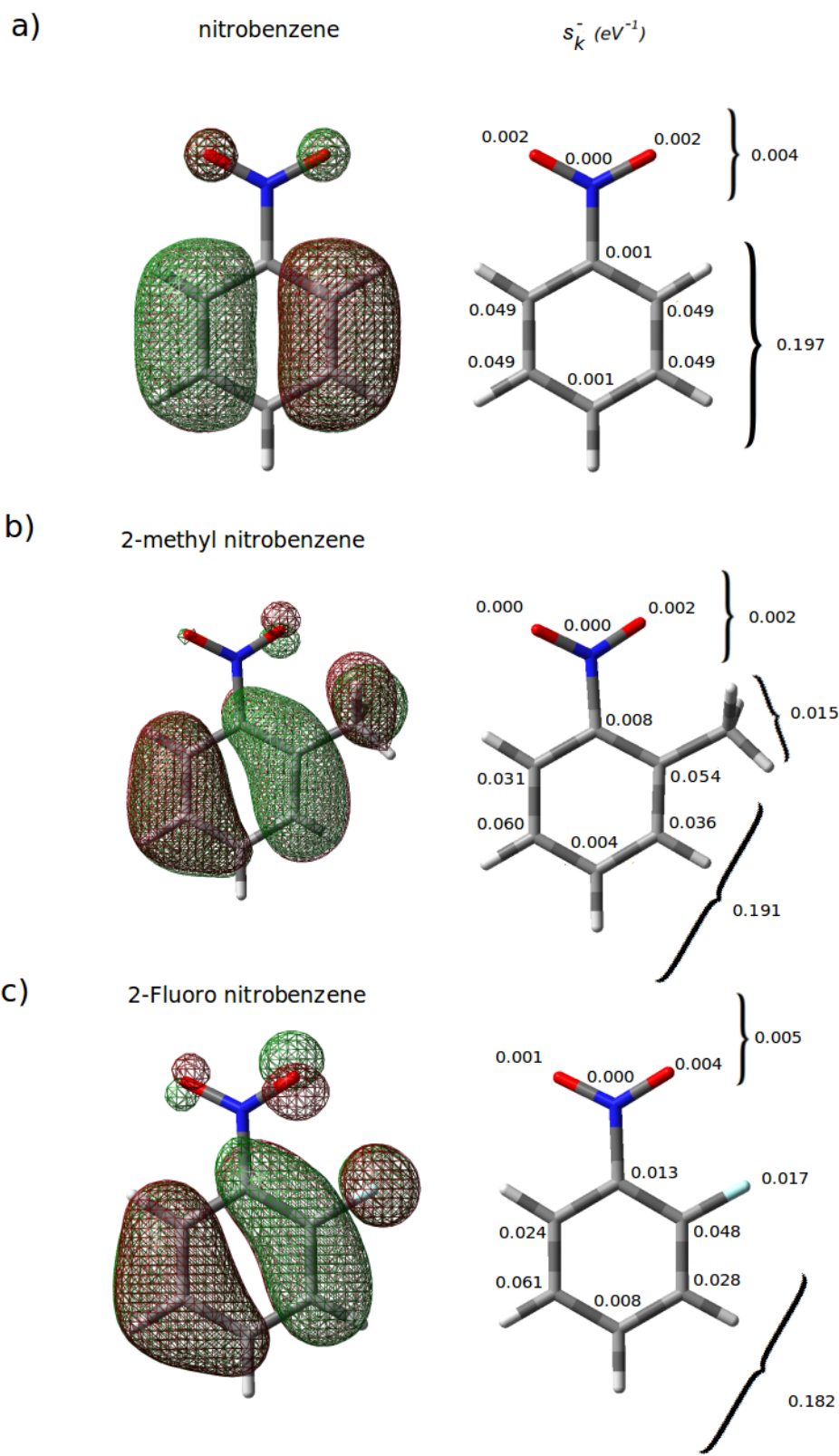

## Acknowledgements

Funding provided by U.S. ARMY Project W911NF−15−1−0632 & Fondecyt Grant 1131003 (F Gonzalez-Nilo); 3110149 (G. Miño-Galaz) & 1120603 (G.Gutierrez.)